\input psfig

%%%
%%%   ftmacros.tex
%%%   Version JF 15/09/93
%%%
%%%%%%%     SELECT YOUR MACHINE NOW!
\newcount\machine
%\global\machine=1   % for macintosh with msxm fonts
\global\machine=2   % for unix with with msam fonts
%%%%%
%%%%%   To produce the ftmacros documentation (for a Macintosh):
%%%%%      1. remove the comment symbol % in the last line of this file
%%%%%               to make the last line \input ftmacrosdoc
%%%%%      2. tex this file
%%%%%      3. put back the % to make the last line of this file
%%%%%                   % \input ftmacrosdoc
%%%%%
%%%%%
%%%%%   To produce the ftmacros documentation (for unix):
%%%%%       1. remove the comment symbol in the first line of the file
%%%%%              ftmacrosdoc.tex to make it \input ftmacros
%%%%%       2. tex ftmacrosdoc
%%%%%
%%%%%%% VARIOUS SETTINGS
\magnification1200
\tolerance=10000
\hsize=15truecm\vsize=23truecm
\parindent=40pt
\mathsurround=0pt
%\multiply\baselineskip by 15\divide \baselineskip by 10
\def\showintremarks{y}  %%%    choose this line to include internal remarks
\long\def\intremark#1{\if\showintremarks y
                 \removelastskip\vskip.1in\hrule\penalty500\vskip.05in\noindent
                 {\bf Internal Remark}\penalty500\hfill\break
                 #1
                 %\hfill\break{\bf End Internal Remark}
                 \vskip.05in\penalty500\hrule\vskip.1in\penalty-500
                 \fi}
%
%     MACRO FOR DRAWING BOXES
%
%     boxes drawn will have rules 1 point wide surrounding a 2 point wide border
%     
%     You can change these measurements by redefining \rulesz 
%     and \bordersz in your TeX source.
%
%     to show a box without the box affecting other spacing set\rulesz=.1pt
%     \bordersz=-\rulesz
%
%  WARNING: removing the %'s below or adding blank spaces will affect box positioning
%
\newdimen\rulesz \newdimen\bordersz \newdimen\boxshift
\rulesz=1pt \bordersz=2pt 
\def\shbox#1{\setbox111=\hbox{#1\hss}%
\boxshift=\dp111\advance\boxshift by \rulesz\advance\boxshift by \bordersz%
\lower\boxshift\vbox{\hrule height\rulesz%   
\hbox{\vrule width\rulesz\kern\bordersz%                   
\vbox{\kern\bordersz\copy111\kern\bordersz}\kern\bordersz%   
\vrule width\rulesz}\hrule height\rulesz}}% 
\def\today{\ifcase\month\or January\or February\or March\or April\or
     May\or June\or July\or August\or September\or October\or November\or
     December\fi\space\number\day, \number\year}
\def\header#1{\rm\nopagenumbers \hfil\underbar{#1} \hfil\today\bigskip
     \headline={\rm\ifnum\pageno>1 #1\hfil\today\hfil Page \folio\else\hfil\fi}}
\def\dst{\displaystyle}

\def\sst{\scriptstyle}

\def\frac#1#2{\dst {#1\over#2}}     % fractions in displaystyle
   % fractions in textstyle    
\def\pmb#1{\setbox0=\hbox{#1}       % generate bold face
     \kern-.025em\copy0\kern-\wd0
     \kern.05em\copy0\kern-\wd0
     \kern-.025em\box0}             %Knuth puts in \raise.0433em before box0  
%
%  Greek letters
%
\def\al{\alpha}
\def\be{\beta}

\def\de{\delta}
\def\ep{\epsilon}

\def\et{\eta}

\def\la{\lambda}

\def\si{\sigma}
\def\ta{\tau}

\def\ps{\psi}

\def\La{\Lambda}
\def\Si{\Sigma}

\def\Om{\Omega}   
%
% Characters
%
\def\0{{\bf 0}}

\def\cS{{\cal S}}

%
%  Vincent's capital roman double letters
%

\def\FF{\hbox to 8.33887pt{\rm I\hskip-1.8pt F}}
\def\NN{\hbox to 9.3111pt{\rm I\hskip-1.8pt N}}
\def\PP{\hbox to 8.61664pt{\rm I\hskip-1.8pt P}}
\def\QQ{\rlap {\raise 0.4ex \hbox{$\scriptscriptstyle |$}}
{\hskip -4.5pt Q}}
\def\RR{\hbox to 9.1722pt{\rm I\hskip-1.8pt R}}

%\def\ZZ{\hbox to 8.2222pt{\rm Z\hskip-4pt \rm Z}} %\def\ZZ{Z\!\!\!Z}
%
%
%      Blackboard characters
%
\font\tenfrak=eufm10\font\sevenfrak=eufm7\font\fivefrak=eufb5
\newfam\frakfam
     \textfont\frakfam=\tenfrak
     \scriptfont\frakfam=\sevenfrak   
     \scriptscriptfont\frakfam=\fivefrak
\def\frak{\fam\frakfam\tenfrak}

\font \tensans                = cmss10
\font \fivesans               = cmss10 at 5pt

\font \sevensans              = cmss10 at 7pt

\font \twelvesans             = cmss12
\font \smallescriptscriptfont = cmr5
\font \smallescriptfont       = cmr5 at 7pt
\font \smalletextfont         = cmr5 at 10pt
\newfam\sansfam
\textfont\sansfam=\tensans\scriptfont\sansfam=\sevensans
                   \scriptscriptfont\sansfam=\fivesans
\def\sans{\fam\sansfam\tensans}
\def\bbbone{{\mathchoice {\rm 1\mskip-4mu l} {\rm 1\mskip-4mu l}    %\bbbone
{\rm 1\mskip-4.5mu l} {\rm 1\mskip-5mu l}}}
\def\bbbc{{\mathchoice {\setbox0=\hbox{$\displaystyle\rm C$}\hbox{\hbox %\bbbc
to0pt{\kern0.4\wd0\vrule height0.9\ht0\hss}\box0}}
{\setbox0=\hbox{$\textstyle\rm C$}\hbox{\hbox
to0pt{\kern0.4\wd0\vrule height0.9\ht0\hss}\box0}}
{\setbox0=\hbox{$\scriptstyle\rm C$}\hbox{\hbox
to0pt{\kern0.4\wd0\vrule height0.9\ht0\hss}\box0}}
{\setbox0=\hbox{$\scriptscriptstyle\rm C$}\hbox{\hbox
to0pt{\kern0.4\wd0\vrule height0.9\ht0\hss}\box0}}}}
\def\bbbe{{\mathchoice {\setbox0=\hbox{\smalletextfont e}\hbox{\raise   %\bbbe
0.1\ht0\hbox to0pt{\kern0.4\wd0\vrule width0.3pt
height0.7\ht0\hss}\box0}}
{\setbox0=\hbox{\smalletextfont e}\hbox{\raise
0.1\ht0\hbox to0pt{\kern0.4\wd0\vrule width0.3pt
height0.7\ht0\hss}\box0}}
{\setbox0=\hbox{\smallescriptfont e}\hbox{\raise
0.1\ht0\hbox to0pt{\kern0.5\wd0\vrule width0.2pt
height0.7\ht0\hss}\box0}}
{\setbox0=\hbox{\smallescriptscriptfont e}\hbox{\raise
0.1\ht0\hbox to0pt{\kern0.4\wd0\vrule width0.2pt
height0.7\ht0\hss}\box0}}}}
\def\bbbq{{\mathchoice {\setbox0=\hbox{$\displaystyle\rm               %\bbbq
Q$}\hbox{\raise
0.15\ht0\hbox to0pt{\kern0.4\wd0\vrule height0.8\ht0\hss}\box0}}
{\setbox0=\hbox{$\textstyle\rm Q$}\hbox{\raise
0.15\ht0\hbox to0pt{\kern0.4\wd0\vrule height0.8\ht0\hss}\box0}}
{\setbox0=\hbox{$\scriptstyle\rm Q$}\hbox{\raise
0.15\ht0\hbox to0pt{\kern0.4\wd0\vrule height0.7\ht0\hss}\box0}}
{\setbox0=\hbox{$\scriptscriptstyle\rm Q$}\hbox{\raise
0.15\ht0\hbox to0pt{\kern0.4\wd0\vrule height0.7\ht0\hss}\box0}}}}
\def\bbbt{{\mathchoice {\setbox0=\hbox{$\displaystyle\rm              %\bbbt
T$}\hbox{\hbox to0pt{\kern0.3\wd0\vrule height0.9\ht0\hss}\box0}}
{\setbox0=\hbox{$\textstyle\rm T$}\hbox{\hbox
to0pt{\kern0.3\wd0\vrule height0.9\ht0\hss}\box0}}
{\setbox0=\hbox{$\scriptstyle\rm T$}\hbox{\hbox
to0pt{\kern0.3\wd0\vrule height0.9\ht0\hss}\box0}}
{\setbox0=\hbox{$\scriptscriptstyle\rm T$}\hbox{\hbox
to0pt{\kern0.3\wd0\vrule height0.9\ht0\hss}\box0}}}}
\def\bbbs{{\mathchoice                                               %\bbbs
{\setbox0=\hbox{$\displaystyle     \rm S$}\hbox{\raise0.5\ht0\hbox
to0pt{\kern0.35\wd0\vrule height0.45\ht0\hss}\hbox
to0pt{\kern0.55\wd0\vrule height0.5\ht0\hss}\box0}}
{\setbox0=\hbox{$\textstyle        \rm S$}\hbox{\raise0.5\ht0\hbox
to0pt{\kern0.35\wd0\vrule height0.45\ht0\hss}\hbox
to0pt{\kern0.55\wd0\vrule height0.5\ht0\hss}\box0}}
{\setbox0=\hbox{$\scriptstyle      \rm S$}\hbox{\raise0.5\ht0\hbox
to0pt{\kern0.35\wd0\vrule height0.45\ht0\hss}\raise0.05\ht0\hbox
to0pt{\kern0.5\wd0\vrule height0.45\ht0\hss}\box0}}
{\setbox0=\hbox{$\scriptscriptstyle\rm S$}\hbox{\raise0.5\ht0\hbox
to0pt{\kern0.4\wd0\vrule height0.45\ht0\hss}\raise0.05\ht0\hbox
to0pt{\kern0.55\wd0\vrule height0.45\ht0\hss}\box0}}}}
\def\bbbz{{\mathchoice {\hbox{$\sans\textstyle Z\kern-0.4em Z$}}       %\bbbz
{\hbox{$\sans\textstyle Z\kern-0.4em Z$}}
{\hbox{$\sans\scriptstyle Z\kern-0.3em Z$}}
{\hbox{$\sans\scriptscriptstyle Z\kern-0.2em Z$}}}}
%
%   Symbols
%

\def\cartprod{\mathop{\lower2pt\hbox{{\twelvesans X}}}}

\def\bbar#1{\setbox0=\hbox{$#1$}%
     \copy0\kern-\wd0
     \raise.0433em\box0}
\def\optbar#1{\vbox{\ialign{##\crcr\hfil${\scriptscriptstyle(}\mkern -1mu
         \vrule height 1.2pt width 3pt depth -.8pt
         {\scriptscriptstyle)}$\hfil\crcr
          \noalign{\kern-1pt\nointerlineskip}$\hfil\displaystyle{#1}\hfil$\crcr}}}
\def\<{\left<}
\def\>{\right>}

\def\sbar#1{\vbox{\ialign{##\crcr
         \noalign{\kern1pt\nointerlineskip}
         \hfil$ \mkern -1mu\vrule height 1.2pt width 3pt depth -.9pt$\hfil\crcr
         \noalign{\kern0pt\nointerlineskip}
                $\hfil{\sst #1}\hfil$\crcr}}}

    \ifnum\machine=1\font\tenmsa=msxm10\font\fivemsa=msxm5\fi
    \ifnum\machine=2\font\tenmsa=msam10\font\fivemsa=msam5\fi 
    \ifnum\machine=1\font\tenmsb=msym10\font\sevenmsb=msym7\font\fivemsb=msym5\fi
    \ifnum\machine=2\font\tenmsb=msbm10\font\sevenmsb=msbm7\font\fivemsb=msbm5\fi      
    \newfam\msafam
     \textfont\msafam=\tenmsa
     \scriptfont\msafam=\fivemsa   %NOTE small scriptsize
     \scriptscriptfont\msafam=\fivemsa
\def\msafamno{\ifcase \msafam 0\or 1\or 2\or 3\or 4\or 5\or 6\or 7\or 8\or
 9\or A\or B\or C\or D\or E\or F\fi}
\newfam\msbfam
     \textfont\msbfam=\tenmsb
     \scriptfont\msbfam=\sevenmsb   
     \scriptscriptfont\msbfam=\fivemsb
\def\msbfamno{\ifcase \msbfam 0\or 1\or 2\or 3\or 4\or 5\or 6\or 7\or 8\or
 9\or A\or B\or C\or D\or E\or F\fi}

\mathchardef\lle="3\msafamno36
\mathchardef\gge="3\msafamno3E
\mathchardef\smsetminus="2\msbfamno72

\mathchardef\evall="2\msafamno16
%\def\amsle{\hbox{\tenmsa \char 54\hss}}
%\def\amsge{\hbox{\tenmsa \char 62\hss}}
%\def\evall{\hbox{\tenmsa  \char 22\hss}}
%\def\sle{\hbox{\sevenmsa \char 54\hss}}
%\def\sge{\hbox{\sevenmsa \char 62\hss}}
%
% Section heading
%

\font \tbfontss               = cmbx5  scaled\magstep1
\font \tbfonts                = cmbx7  scaled\magstep1
\font \tbfontt                = cmbx10 scaled\magstep1
\font \tasys                  = cmex10 scaled\magstep1
\font \tamsss                 = cmmib10 scaled 833
\font \tamss                  = cmmib10
\font \tams                   = cmmib10 scaled\magstep1
\font \tbst                   = cmsy10 scaled\magstep1
\font \tbsss                  = cmsy5  scaled\magstep1
\font \tbss                   = cmsy7  scaled\magstep1

\def\section#1#2{%
      \vskip25pt plus 4pt minus4pt
     \bgroup
 \textfont0=\tbfontt \scriptfont0=\tbfonts \scriptscriptfont0=\tbfontss
 \textfont1=\tams \scriptfont1=\tamss \scriptscriptfont1=\tamsss
 \textfont2=\tbst \scriptfont2=\tbss \scriptscriptfont2=\tbsss
 \textfont3=\tasys \scriptfont3=\tenex \scriptscriptfont3=\tenex
     \baselineskip=16pt
     \lineskip=0pt
     \rightskip 0pt plus 6em
     \pretolerance=10000
     \tbfontt
     \setbox0=\vbox{\vskip25pt
     \noindent
     \if!#1!\ignorespaces#2
     \else\ignorespaces#1\unskip\enspace\ignorespaces#2\fi
     \vskip15pt}%
     \dimen0=\pagetotal
     \ifdim\dimen0<\pagegoal
     \dimen0=\ht0\advance\dimen0 by\dp0\advance\dimen0 by
     4\normalbaselineskip
     \advance\dimen0 by\pagetotal
     \advance\dimen0 by-\pageshrink
     \ifdim\dimen0>\pagegoal\vfill\eject\fi\fi
     \noindent
     \if!#1!\ignorespaces#2
     \else\ignorespaces#1\unskip\enspace\ignorespaces#2\fi
     \egroup
     \vskip12.5pt plus4pt minus4pt
     \ignorespaces}
%
%  Theorem etc.
%
\def\newenvironment#1#2#3#4{\long\def#1##1##2{\removelastskip\penalty-100
\vskip\baselineskip\noindent{#3#2\if!##1!.\else\unskip\ \ignorespaces
##1\unskip\fi\ }{#4\ignorespaces##2\vskip\baselineskip}}}
\newenvironment\lemma{Lemma}{\bf}{\it}
\newenvironment\proposition{Proposition}{\bf}{\it}
\newenvironment\theorem{Theorem}{\bf}{\it}
\newenvironment\corollary{Corollary}{\bf}{\it}
\newenvironment\example{Example}{\it}{\rm}
\newenvironment\problem{Problem}{\bf}{\rm}
\newenvironment\definition{Definition}{\bf}{\rm}
\newenvironment\remark{Remark}{\bf}{\it}

%
%    Proof, qed, equation
%
\long\def\proof#1{\removelastskip\penalty-100\vskip\baselineskip\noindent{\bf
            Proof\if!#1!\else\ \ignorespaces#1\fi:\ }\ \ \ignorespaces}
\long\def\prf{\removelastskip\penalty-100\vskip\baselineskip\noindent{\bf
            Proof:\ }\ \ \ignorespaces}
\def\sq{\hbox{\rlap{$\sqcap$}$\sqcup$}}
\def\qed{\ifmmode\sq\else{\unskip\nobreak\hfil
           \penalty50\hskip1em\null\nobreak\hfil\sq
           \parfillskip=0pt\finalhyphendemerits=0\endgraf}\fi}

\def\+{\!+\!}
\def\-{\!-\!}
\def\={\ =\ }

\def\eval#1{\big|\lower4pt\hbox{$\displaystyle\sst #1$}}

\def\cV{{\cal V}}

\def\Pn{{{\cal P}_n}}

\def\gF{{\frak F}}
\def\gT{{\frak T}}

\centerline{}
\vskip 2cm
\centerline{\bf Explicit Fermionic Tree Expansions}
\vskip 2cm

\centerline{A. Abdesselam\footnote*{Presently at
the University of British Columbia, Department of Mathematics, Vancouver B.C.
Canada V6T 1Z2.}
\& V. Rivasseau}

\centerline{Centre de Physique Th\'eorique, CNRS UPR 14} 
\centerline{Ecole Polytechnique}
\centerline{99128 Palaiseau Cedex, FRANCE}

\vskip 1cm
\medskip
\noindent{\bf Abstract}
We express connected Fermionic Green's
functions in terms of two totally explicit tree formulas.
The simplest and most symmetric formula, the Brydges-Kennedy
formula is compatible with Gram's inequality.
The second one, the rooted formula of [AR1], respects even better
the antisymmetric structure of determinants, and allows
the direct  comparison of rows and columns which correspond to
the mathematical implementation  in Grassmann integrals
of the Pauli principle. To illustrate the power of these formulas,
we give a ``three lines proof''
that the radius of convergence of the Gross-Neveu theory
with cutoff is independent of the number of colors, using
either one or the other of these formulas.

\medskip

\vskip1cm
\noindent{\bf I) Introduction}
\medskip
Perturbation theory for fermion systems is often said to converge,
whether for boson systems it is said to diverge. 
But what does this mean exactly?
Unnormalized Fermionic perturbation series with cutoffs
are not only convergent but {\it entire}, whether Bosonic perturbation 
series with cutoffs have zero radius of convergence. 
But this is of little use since
in actual computations, especially renormalization group
computations, one needs to compute {connected functions}. 
It was not obvious until recently how to compute these
connected functions in a way which preserves
the algebraic cancellations of the Pauli
principle to obtain convergent power series 
with accurate and easy estimates on the convergence radius,
without using the heavy techniques of cluster and Mayer expansions.

The first solution to this problem which does not require
cluster and Mayer expansions is in [FMRT]
where some inductive way of computing such Fermionic connected
functions is given and applied to the uniform
convergence of two-dimensional many fermion systems in a slice.
However in the corresponding computation 
the expansions steps were intertwinned with 
constructions of layers of some tree, and the final outcome was 
therefore not totally explicit yet.

In [AR1] we developed two explicit tree formulas which are
especially suited for cluster and Mayer expansions. 
Shortly thereafter we made the 
remark that the second formula leads to a completely
explicit version of the inductive cluster expansion used in [FMRT] and [FKLT].
More recently we realized that the first formula is also compatible
with Gram's inequality, and can therefore also be used
in fermionic expansions. It is possible to develop a continuous
renormalization group formalism for Fermions using this formula [DR], a 
problem which has attracted attention recently [S].

We also mention that an other (closely related)
version of this kind of expansions has been introduced recently,
using Gram's inequality [FKT]. It 
relies on a resolvent expansion rather than an explicit tree formula.

\medskip
\noindent{\bf II. Tree formulas and Grassmann integrals}
\medskip

In this section we recall the notations and the ``Forest Formulas''  of [AR1]
and apply them to Grassmann (fermionic) integrals.

Let $n\ge 1$  be an integer, 
$I_n=\{1,\ldots,n\}$, 
$\Pn=\bigl\{\{i,j\}/i,j\in I_n,i\ne j\bigr\}$ (the set of unordered pairs
in $I_n$).
An element $l$ of $\Pn$ will be called a {\sl link}, a subset of $\Pn
$, a {\sl graph}. A graph $\gF=\{l_1,\ldots,l_{\tau}\}$ containing no loops,
i.e.{} no subset $\bigl\{\{i_1,i_2\},\{i_2,i_3\},\ldots,\{i_k,i_1\}\bigr\}$
with $k\ge 3$ elements, is called a forest.

A Forest formula is a Taylor formula with
integral remainder, which expands a quantity such as $
\exp \biggl({\sum_{l\in \Pn} u_l}\biggr)$ to search for the explicit
presence or absence of links $u_{l}$. Taylor formulas with remainders
in general are provided with a ``stopping rule'' 
and forests formulas stop at the level of 
connected sets. This means that
two points which are joined by a link are treated as a 
single block.
(More sophisticated formulas with higher stopping rules
are useful for higher particle irreducibility analysis, 
or renormalization group computations
but are no longer forests formulas in the strict sense [AR2]). Any such forest
formula contains a ``weakening factor'' $w$ for the links which remain
underived. Many different such forests formulas exist, with different 
rules for $w$. Two of them were identified as the most natural ones
in [AR1], corresponding to two different logics: in one of them,
the Brydges-Kennedy formula,
the forest grows in the most symmetric and random way, and in the
other, the ``rooted formula'' it grows layers by layers from
a preferred root.  

A forest is a union of disconnected trees $\gT$, the supports of which are 
disjoint subsets of $I_{n}$ called the 
{{\sl connected components} or {\sl clusters} of $\gF$. 

The Brydges-Kennedy formula is then 
\medskip
\noindent{\bf Brydges-Kennedy Forest Formula}{\it
$$
\exp \biggl({\sum_{l\in \Pn} u_l}\biggr)=\sum_{\gF=\{l_1,\ldots,l_{\tau}\}
\atop{\rm u-forest}}\biggl(\prod_{\nu=1}^{\tau}\int_0^1dw_{l_{\nu}}\biggr)
\biggl(\prod_{\nu =1}^{\tau}u_{l_{\nu}}\biggr)\exp\bigl(\sum_{l\in \Pn}
w_l^{\gF,BK}({\bf w}).u_{l}\bigr)\ ,
\eqno({\rm II.1})
$$
where the summation extends over all possible lengths 
$\tau$ of $\gF$, including
$\tau=0$ hence the empty forest.
To each link of $\gF$ is attached a variable of integration $w_l$; and
$w_{\{ij\}}^{\gF,BK}({\bf w})=\inf \bigl\{w_l,l\in L_{\gF}\{ij\}\bigr\}$ where
 $L_{\gF}\{ij\}$ is the unique path in the forest $\gF$ connecting i to j.
If no such path exists, by convention $w_{\{ij\}}^{\gF,BK}({\bf w})=0$.}

The rooted formula is absolutely identical, but with a different
rule for the weakening factor $w$, now called 
$w_{\{ij\}}^{\gF,R}({\bf w})$. It is a less symmetrical formula since we 
have to give a rule for choosing a root in each cluster.
For each non empty subset or cluster $C$ of $I_n$,
choose $r_C$, for instance the least element in the natural ordering
of $I_n=\{1,\ldots,n\}$, to be the root of all the trees with support $C$
that appear in the following expansion.
Now if $i$ is in some tree $\gT$ with support $C$ we call the {\sl height}
of $i$ the number of links in the unique path of the tree $\gT$ that goes from
$i$ to the root $r_C$. We denote it by $l^\gT(i)$. The set of points $i$
with a fixed height $k$ is called the $k$-th {\sl layer} of the tree.
The Rooted Forest Formula is then:
\medskip
\noindent{\bf Rooted Forest Formula}{\it
$$
\exp \biggl({\sum_{l\in \Pn} u_l}\biggr)=\sum_{\gF=\{l_1,\ldots,l_{\tau}\}
\atop{\rm u-forest}}\biggl(\prod_{\nu=1}^{\tau}\int_0^1dw_{l_{\nu}}\biggr)
\biggl(\prod_{\nu =1}^{\tau}u_{l_{\nu}}\biggr)\exp\bigl(\sum_{l\in \Pn}
w_l^{\gF,R}({\bf w}).u_{l}\bigr)\ ,
\eqno({\rm II.2})
$$
where the only difference with (II.1) lies in the definition of 
$w_{\{ij\}}^{\gF,R}({\bf w})$ which is different from the one of 
$w_{\{ij\}}^{\gF,BK}({\bf w})=0$. We define still 
$w_{\{ij\}}^{\gF,R}({\bf w})=0$ if $i$ and $j$ are not connected by the $\gF$.
If $i$ and $j$ fall in the support $C$ of the same tree $\gT$ of $\gF$ then

$w_{\{ij\}}^{\gF,R}({\bf w})=0$\ \ if\ \ $|l^\gT(i)-l^\gT(j)|\ge2$\ \ 
($i$ and $j$ in distant layers)

$w_{\{ij\}}^{\gF,R}({\bf w})=1$\ \ if\ \ $l^\gT(i)=l^\gT(j)$\ \ 
($i$ and $j$ in the same
layer)

$w_{\{ij\}}^\gF({\bf w})=w_{\{ii'\}}$\ \ if\ \ $l^\gT(i)-1=l^\gT(j)=l^\gT(i')$,
and $\{ii'\}\in\gT$. ($i$ and $j$ in neighboring layers, $i'$ is then unique
and is called the ancestor of $i$ in $\gT$).
}
\medskip

Proofs of these formulas are given in [AR1].

To any such algebraic formula, there 
is an associated ``Taylor'' tree formula,
in which $u_{l}$ is interpreted not as a real number but as a differential 
operator $u_{l}= {\partial \over \partial x_{l}}$ [AR1]. 
Let us 
recall these associated
Taylor formulas.
Let $\cS$ be the space of smooth functions from
$\RR^\Pn$ to an arbitrary Banach space $\cV$.
An element of $\RR^\Pn$ will be generally denoted by ${\bf x}={(x_l)}_
{l\in\Pn}$. The vector with all entries equal to $1$ will be denoted by
$\bbbone$. Applied to an element $H$ of $\cS$, the
Taylor Rooted Forest formula takes the form:

\eject
\noindent{\bf The Taylor forest formulas}
\medskip
{\it 
$$
H(\bbbone)=
\sum_{\gF\ {\rm u-forest}}
\biggl(\prod_{l\in\gF}\int_0^1 dw_l\biggr)
\biggl(\Bigl(\prod_{l\in\gF}
{\partial \over {\partial x_l}}\Bigr)H\biggr)
\bigl(X_\gF^{BK \ {\rm or}\  R}({\bf w})\bigr) \ \ .
\eqno({\rm II.3})
$$
\noindent Here $X_\gF^{BK \ {\rm or}\  R}({\bf w})$ 
is the vector ${(x_l)}_{l\in\Pn}$
of $\RR^\Pn$ defined by $x_l=w_l^{\gF.BK \ {\rm or}\  R}({\bf w})$, 
which is the value at which we evaluate the
derivative of $H$. The symbol $BK \ {\rm or}\  R$ means that the formula
is true using either the Brydges-Kennedy or the rooted
weakening factor $w$.}

Again the proof of (II.3) is in [AR1]
(but we recall that it is a rather trivial consequence of
(II.1-2) applied to $u_{l}= {\partial \over \partial x_{l}}$).

Any forest is a union of connected trees. Therefore
any forest formula has an associated tree formula for its
connected components. And therefore, at least formally,
any forest formula solves the
problem of computing normalized correlation functions. Indeed
applying the forest formula to the functional integral for
the unnormalized functions,   
the connected functions
are simply given by the connected pieces of the forest
formula, hence by the corresponding tree formula. It is in this sense
that forest formulas exactly solve the well known snag that 
makes connected functions difficult to compute. This snag is that since
typically there are many trees in a graph, one does not know ``which one to
choose'' when one tries
to compute connected functions in the (desirable) form of tree sums. 
Any forest formula gives
a particular answer to that problem. It tells us exactly by how much
our pondered ``tree choice'' has ``weakened'' 
the remaining loop lines: just by the
weakening factor $w$.

\medskip
\medskip
\centerline{\hskip2cm\hbox{\psfig{figure=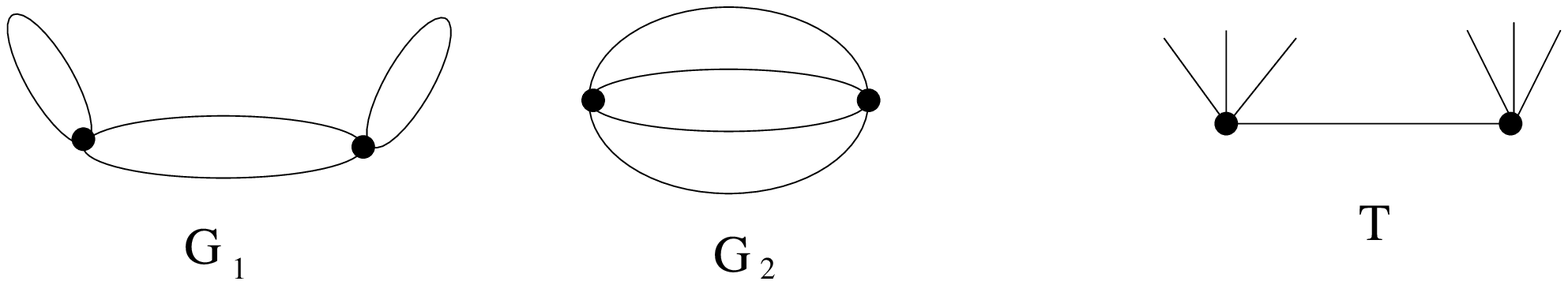,height=2.9cm,width=14cm}}}

\medskip
\centerline{Figure 1}
\medskip

To illustrate this by the simplest possible example, let us consider the
second order connected graphs of the pressure for the $\phi^{4}$
theory in dimension 0 with propagator 1, which simply
counts multiplicities of connected graphs. There are two
second order connected graphs, 
pictured in Fig 1. To count contractions correctly
it is convenient
to label all fields or half lines. The first graph $G_{1}$
corresponds to 2.$C_{4}^{2}$.$C_{4}^{2}$ = 72  contractions, and
has therefore weight 72. 
The second graph has weight 4!=24.
Now to select a tree in this case is to select the trivial trees
$T$ and there are $4^{2}=16$ such trees. If the same measure was used for
the contractions of the three loop lines, we would find a weight 
16.$C_{3}^{2}$.$C_{3}^{2}$ =144 for $G_{1}$, and a weight 16.3! = 96
for  $G_{2}$. This illustrates the overcounting problem of
choosing a tree, and we see that this overcounting varies with the graphs.
Now we see how the weakening of the loop lines corrects precisely this
defect: with the weakening parameters $w^{BK}$ or $w^{R}$, the graph
$G_{1}$ is weakened by a factor $\int_{0}^{1}wdw=1/2$ and the graph
$G_{2}$ is weakened by a factor $\int_{0}^{1}w^{3}dw=1/4$ so that
correct weights are recovered! The reader is invited to try 
higher order cases to see the distinction between 
$w^{BK}$ and $w^{R}$ appear.

It remains to see which formulas
can be used in the constructive sense and for which theories.
It is well known that the Brydges-Kennedy Formula 
[BK][AR1], when applied to interpolate 
a symmetric positive matrix, preserves positivity. 
Indeed $w_{ij}^{\gF,BK}(\bf w)$ completed as a symmetric matrix
by the definition $w_{ii}^{\gF,BK}({\bf w}) =1 \forall i$
is a positive matrix, since it is a convex combination of block matrices
with 1 everywhere [AR1].
 Therefore if $C$ is a positive matrix (like a bosonic covariance)
the matrix $w_{ij}^{\gF,BK}({\bf w}) C(x_{i},x_{j})$, being the Hadamard
product of two positive matrices, remains positive for any values
of the tree parameters $w$. This means that it is
suited to the construction of bosonic models 
([B][AR2] and references therein). But this positivity
property also allows Gram's estimates for fermionic theories:
\medskip
\noindent{\bf Lemma}
\medskip
{\it
Let ${\cal A}=(a_{\al\be})_{\al,\be}$
be a Gram matrix: $a_{\al\be}=<f_\al,g_\be>$ for some inner product
$<.,.>$.
Suppose each of the indices $\al$ and $\be$ is of the form $(i,\si)$
where the first index $i$, $1\le i\le n$, has the same range as the indices
of the positive matrix $(w_{ij}^{\gF,BK}({\bf w}))_{ij}$, and $\si$
runs through some other index set $\Si$.

Let ${\cal B}=(b_{\al\be})_{\al,\be}$ be the matrix with entries
$b_{(i,\si)(j,\ta)}=w_{ij}^{\gF,BK}({\bf w})
.<f_{(i,\si)},g_{(j,\ta)}>
$
and let $(b_{\al\be})_{\al\in A,\be\in B}$
be some square matrix extracted from $\cal B$, then for any ${\bf w}$
we have the Gram inequality:
$$
|det(b_{\al\be})_{\al\in A,\be\in B}|\le
\prod_{\al\in A} ||f_\al||
\prod_{\be\in B} ||g_\be||
\ \ .
\eqno{(\rm II.4)}
$$
}

\prf  Indeed we can take the symmetric square root $v$ of the 
positive matrix $w^{\gF,BK}$ 
so that $w^{\gF,BK}_{ij} = \sum_{k=1}^n v_{ik}v_{kj} $. 
Let us denote the components of the vectors $f$ and $g$, in an orthonormal
basis for the scalar product $<.,.>$ with $q$ elements, by
$f_{(i,\si)}^m$ and $g_{(j,\ta)}^m$, $1\le m\le q$. (Indeed even if 
the initial Hilbert space is infinite dimensional, the problem is obviously 
restricted to the
finite dimensional subspace generated by the finite set of vectors 
$f$ and $g$).
We then define the tensorized vectors $F_{(i,\si)}$ and $G_{(j,\ta)}$
with components $F_{(i,\si)}^{k m}=v_{i k}f_{(i,\si)}^m$ and
$G_{(j,\ta)}^{k m}=v_{j k}g_{(j,\ta)}^m$ where $1\le k\le n$ and
$1\le m\le q$.
Now considering the tensor scalar product $<.,.>_T$ we have
$$
<F_{(i,\si)},G_{(j,\ta)}>_T=\sum_{k=1}^n\sum_{m=1}^q
v_{ik}v_{jk}f_{(i,\si)}^m g_{(j,\ta)}^m
=b_{(i,\si)(j,\ta)}
\ \ .\eqno{(\rm II.5)}
$$
By Gram's inequality using the $<.,.>_T$ scalar product we get
$$
|det(b_{\al\be})_{\al\in A,\be\in B})|\le
\prod_{\al\in A} {||f_\al||}_T
\prod_{\be\in B} {||g_\be||}_T
\eqno{(\rm II.6)}
$$
but ${||F_{(i,\si)}||}_T^2=\sum_{k=1}^n\sum_{m=1}^q(F_{(i,\si)}^{km})^2=
\sum_{k=1}^n\sum_{m=1}^q v_{ik}^2(f_{(i,\si)}^m)^2
=w_{ii}\sum_{m=1}^q(f_{(i,\si)}^m)^2={||f_{(i,\si)}||}^2$,
since $w_{ii}=1$ for any $i$, $1\le i\le n$.
\qed

Let us now
apply the Brydges-Kennedy formula 
to the computation of the connected functions of a Fermionic theory.
The corresponding Grassmann integral
is:

$${1\over Z}\int d\mu_{C}(\ps, \bar \ps )
P(\bar\psi_{a}, \psi_{a} )e^{S(\bar\ps_{a}, \ps_{a})}
\eqno({\rm II.7})
$$
where $C$ is the covariance or propagator,
$P$ is a particular monomial (set of external fields)
and $S$ is some general action. We take as simplest example
the massive Gross-Neveu model with cutoff, for which the 
action is local and quartic in a certain number of 
Fermionic fields. These fields are two-components
because of the spin.
In a finite box $\Lambda$ the action is
$$S_{\Lambda}= {\lambda \over N}\int_{\Lambda} dx 
(\sum_{a}\bar\ps_{a}(x)\ps_{a}(x))^{2}
\eqno({\rm II.8})
$$
where $a$ runs over some finite
set of $N$ ``color'' indices. 
$\lambda$ is the  coupling constant.
The covariance $C$ is massive and has an ultraviolet
cutoff, hence in Fourier space it is for instance
$\et(p)/(\not p +m)$  where $\et$ is a cutoff function on large momenta.
We only need to know that $C$ is diagonal in color space
$ C(x, a; y, a') = 0 $ for $a \ne a'$, and that it can be decomposed
as 
$$
C(x,y)= \int_{\RR^{d}} D(x,t) \bar E(t,y)dt
\eqno{(\rm II.9a)}
$$
with
$$ |C(x,y) |\le K {1 \over (1 + |x-y|)^{p}} \eqno({\rm II.9b})
$$
$$ \int_{\RR^{d}} |D(x,t)|^{2}d^{d}t \le K \quad ;
\quad \int_{\RR^{d}} |E(x,t)|^{2} d^{d}t\le K \eqno({\rm II.9c})
$$
for some constant $K$.

This decomposition amounts
roughly to defining
square roots of the covariance in momentum
space and prove their square integrability. It is usually  
easy for any reasonable cutoff model. For instance if $\et$ 
is a positive function, we can define $D$ in Fourier space
as ${\et^{1/2}(p) \over (p^{2} + m^{2})^{1/4}}$ 
and $E$ as ${(-\not p + m) \et^{1/2}(p) \over (p^{2} + m^{2})^{3/4}}$.

It is the Fermionic covariance or propagator $C= <\ps_{a}(x)
\bar\ps_{a}(y)>$ which is interpolated with the forest formulas. 
We obtain, for instance
for the pressure that the formal power series in the coupling constant is
a sum over trees on $\{1,....,n\}$, with 1, a distinguished
vertex sitting at the origin to break translation invariance
(similar formulas with external fields of course
exist for the connected functions):

\medskip
\noindent{\bf Fermionic Tree Expansion}
{\it
$$
p= \lim_{\La \to \infty} {1\over |\La|} Log Z(\Lambda) =  
\lim_{\La \to \infty} {1\over |\La|}\bigl(
\int d\mu_{C }(\ps, \bar \ps )e^{S_{\Lambda}(\bar\ps_{a}, \ps_{a})}\bigr)
$$
$$
= \sum_{n=0}^{\infty}(\la^{n}/N^{n}n!)
\sum_{a_1,\ldots,a_n,b_1,\ldots,b_n=1}^N
\sum_{\gT}\sum_{\Om}\ep(\gT,\Om)
\bigl(
\prod_{l\in \gT}\int_{0}^{1} dw_{l} \bigr)
$$ 
$$
\int_{\RR^{nd}} dx_{1}...dx_{n} \de(x_{1}=0)
\prod_{l\in \gT}\bigl(C(x_{i(l)},x_{j(l)})\de_l\bigr)
\times
det(b_{\al\be})_{\al\in A,\be\in B}
\ \ .\eqno{(\rm II.10)}
$$
The sum over the $a_i$'s and $b_i$'s 
are over the colors of the fields and antifields of the vertices obtained by
expanding the interaction and of the form: 
$$
{\bar\ps}_{a_i}(x_i)\ps_{a_i}(x_i)
{\bar\ps}_{b_i}(x_i)\ps_{b_i}(x_i)
\eqno{(\rm II.11)}
$$
with $1\le i\le n$.
The sum over $\gT$ is over all trees which connect together
the $n$ vertices at $x_1,\ldots,x_n$.
The sum over $\Om$ is over the compatible ways of realising the bonds
$l=\{i,j\}\in\gT$ as contractions of a $\ps$ and ${\bar \ps}$ between
the vertices $i$ and $j$ (compatible means that we do not contract twice
the same field or antifield).
A priori there are 8 choices of contraction for each $l=\{i,j\}$,
including the choice in direction for the arrow. $\ep(\gT,\Om)$ is a
sign we will express later.
For any $l\in\gT$, $i(l)\in\{1,\ldots,n\}$ labels the vertex where the field,
contracted by the procedure $\Om$ concerning the link $l$, was chosen.
Likewise $j(l)$ is the label for the vertex containing the contracted
antifield.
$\de_l$ is 1 if the colors (among $a_1,\ldots,a_n,b_1,\ldots,b_n$)
of the field and antifield contracted by $l$ are the same and else is 0.
Finally the matrix $(b_{\al\be})_{\al,\be}$ is defined in the following
manner.

The row indices $\al$ label the $2n$ fields produced by the $n$ vertices,
so that $\al=(i,\si)$ with $1\le i\le n$ and $\si$ takes two values 1 or 2
to indicate whether the field is the second or the fourth factor in (II.11)
respectively.
The column indices $\be$ label in the same way the $2n$ antifields,
so that $\be=(j,\ta)$ with $1\le j\le n$ and $\ta=1$ or 2 according to whether
the antifield is the first or the third factor in (II.11) respectively.
The $\al$'s and $\be$'s are ordered lexicographically. We denote by
$c(i,\si)$ the color of the field labeled by $(i,\si)$ that is
$a_i$, if $\si=1$, and $b_i$ if $\si=2$. We introduce the similar notation
${\bar c}(j,\ta)$ for the color of an antifield.
Now 
$$
b_{(i,\si)(j,\ta)}=w_{ij}^{\gT,BK}({\bf w})C(x_i,x_j)
\de(c(i,\si),{\bar c}(j,\ta))
\ \ .\eqno{(\rm II.12)}
$$
Finally each time a field $(i,\si)$ is contracted by $\Om$
the corresponding row is deleted from the $2n\times 2n$ matrix $(b_{\al\be})$.
Likewise, for any contracted antifield the corresponding column is erased.
$A$ and $B$ denote respectively the set of remaining rows and the set
of remaining columns. The minor determinant featuring in formula (II.10)
is now $det(b_{\al\be})_{\al\in A,\be\in B}$
which is $(n+1)\times(n+1)$.
Indeed for each of the $n-1$ links of $\gT$, a row and a column are erased.

Although it is not important for the bounds, we indicate the rule for
computing the sign $\ep(\gT,\Om)$.
Let $(1,1)\le\al_1<\cdots<\al_{n-1}\le(n,2)$
and $(1,1)\le\be_1<\cdots<\be_{n-1}\le(n,2)$
be the erased rows and columns respectively.
Suppose that $\al_r$ was contracted by $\gT$ and $\Om$ together with
$\be_{\pi(r)}$, for $1\le r\le n-1$.
Clearly $\pi$ is a permutation of $\{1,\ldots,n-1\}$. Then
$$
\ep(\gT,\Om)=(-1)^{\si(\al_1)+\cdots+\si(\al_{n-1})+
\ta(\be_1)+\cdots+\ta(\be_{n-1})}
\ep(\pi)
\eqno{(\rm II.13)}
$$
where $\ep(\pi)$ is the signature of $\pi$ and for any $\al=(i,\si)$
we introduced the notation
$i(\al)=i$ and $\si(\al)=\si$ and similarly for any $\be=(j,\ta)$,
we write $j(\be)=j$ and $\ta(\be)=\ta$.
}

\prf 
The proof of the previous expansion although a bit tedious is straightforward.
First expand $\exp S_\La({\bar\ps},\ps)$. Each term will involve a Berezin
integral
$$
\int d\mu_C(\ps,{\bar\ps})\prod_{i=1}^n
\bigl(
{\bar\ps}_{a_i}(x_i)\ps_{a_i}(x_i){\bar\ps}_{b_i}(x_i)\ps_{b_i}(x_i)
\bigr)
\eqno{(\rm II.14)}
$$
which is a $2n\times 2n$ determinant like the one of the matrix $(b_{\al\be})$.
Then in order to use (II.3) we introduce for any pair $\{i,j\}\in{\cal P}_n$
a weakening factor $x_{\{i,j\}}$ multiplying each of the eight entries
involving both vertices $i$ and $j$.
The output of (II.3) is a forst formula when applied to $Z(\La)$.
However the amplitudes corresponding to each connected tree in the forest
factorize (actually, to check that, one needs to be careful and compute
$\ep(\gT,\Om)$).
Taking the sum over trees instead of forests simply computes $\log Z(\La)$
instead of $Z(\La)$. Finally $\de(x_1=0)$ accounts for the division by
the volume $|\La|$ to get the pressure as an intensive quantity.
\qed

\medskip

Let us now use this fermionic tree formulas for proofs of convergence.

\medskip
\noindent{\bf III) Convergence of the 
tree formulas}
\medskip

A typical constructive result for this Gross-Neveu model with cutoff 
is to prove:

\medskip
\noindent{\bf Theorem}
\medskip

{\it The pressure and the connected functions of the cut-off Gross-Neveu
model are analytic in $\lambda$ in a disk of radius $R$
independent of $N$.
}
\medskip

\prf 
The determinant $det(b_{\al\be})_{\al\in A,\be\in B}$
is diagonal in the colors of the involved fields and antifields.
Each of the block determinants falls in the category described by the lemma,
and thus the complete determinant is bounded by $K^{n+1}$ thanks to the
decomposition (II.9a).
The spatial integrals
are bounded using (II.9b) for the propagators corresponding to the links
in $\gT$, by $K^{n-1}$. The sum over $\Om$ is bounded by $8^{n-1}$
(the number of compatible contractions is actually equal to
$8\times 12^{n_4}\times 10^{n_3}\times 6^{n_2}$
where $n_p$ is the number of vertices of the
tree $\gT$ with coordination number $p$).
The sum over colors is bounded by $N^{n+1}$, indeed once $\gT$ and
$\Om$ are known, the circulation of color indices is determined.
The attribution of color indices costs $N^2$ at the first vertex and by
induction a factor $N$ for each of the remaining vertices of the tree.
Indeed climbing inductively into the tree layer by layer, at every vertex
there is one color already fixed by the line joining the vertex 
to the root, hence one remaining color to fix, except for the root, 
for which two colors have to be fixed.
The number of $\gT$'s is bounded by $n^{n-2}$, by Cayley's theorem.
Finally the $n$-th term of the series is bounded by
$(|\la|^n/N^n n!)K^{n+1}K^{n-1}8^{n-1}N^{n+1}n^{n-2}$
and the radius of convergence is thus at least $1/8 e K^2$.
\qed

This is perhaps the shortest and most transparent proof
of constructive theory yet! Remark in particular that 
it does not require any discretization of space, lattice of cubes,
cluster or Mayer expansion, all features which are necessary
for bosonic theories. This ``three lines'' 
treatment of the theory with cutoff can presumably be
extended into a ``three pages'' treatment
of for instance the Gross-Neveu theory with renormalization 
in two dimensions [DR]. Such
a treatment with no discretization of space,
no discrete slicing of momenta, and
continuous instead of discrete renormalization group
equations makes the constructive version of these fermionic
theories almost identical to their perturbative
version (and no longer more difficult).

The theorem above is interesting not only for the analysis of the Gross-Neveu
model but also for that of the two-dimensional
interacting Fermions considered in [FMRT] or [FKLT]. In this latter case,
the ``colors'' correspond to angular sectors on the Fermi
sphere and the factor $1/N$ in the coupling is provided by power counting.

We now give a second and longer proof of the theorem
using the rooted formula. This is
worth the trouble since the weakening factor
in the rooted formula completely factorizes out of the determinant.
This second formula may therefore be useful 
in problems for which Gram's inequality is not 
applicable and the method of ``comparison of rows and columns''
of [IM] or [FMRT] has to be used. 

We return to formula (II.10) but use the rooted weakening
factor. To bound the determinant, we
need to introduce simply one further subtle modification. 
We need  a further sum which decides
for each field or antifield whether it contracts or not to the
level in the tree immediately below it.  

The only change in formula (II.10) is that the loop determinant has entries
$$
b_{\al\be}=w_{i(\al)j(\be)}^{\gT,R}({\bf w}) C(x_{i(\al)},x_{j(\be)})
\de(c(\al),{\bar c}(\be))
\eqno{(\rm III.1)}
$$
with $\al\in A$, $\be\in B$.
Any of these entries we write as:
$$
b_{\al\be}=\sum_{u(\al)\in\{\sharp,\natural,\flat\}}
\sum_{v(\be)\in\{\sharp,\natural,\flat\}}b_{\al\be}^{u(\al)v(\be)}
\eqno{(\rm III.2)}
$$
with the convention that
$$
b_{\al\be}^{\natural\natural}=b_{\al\be}\de(l^\gT(i(\al)),l^\gT(j(\be)))
\eqno{(\rm III.3a)}
$$
$$
b_{\al\be}^{\sharp\flat}=b_{\al\be}\de(l^\gT(i(\al))+1,l^\gT(j(\be)))
\eqno{(\rm III.3b)}
$$
$$
b_{\al\be}^{\flat\sharp}=b_{\al\be}\de(l^\gT(i(\al))-1,l^\gT(j(\be)))
\eqno{(\rm III.3c)}
$$
and $b_{\al\be}^{u(\al)v(\be)}=0$
whenever $(u(\al),v(\be))=(\natural, \sharp),
(\sharp, \natural), (\flat, \natural), (\natural, \flat), (\flat, \flat)$
or $ (\sharp, \sharp)$.

Then we expand the determinant $det(b_{\al\be})_{\al\in A,\be\in B}$
by multilinearity with respect to rows,
having for any $\al\in A$, a sum over $u(\al)\in\{\sharp,\natural,\flat\}$.
Then we do the same thing for columns and we get a sum over
$v(\be)\in\{\sharp,\natural,\flat\}$, for any $\be\in B$.
We then obtain an expansion into $3^{2n+2}$ determinants,
of the form $det(b_{\al\be}^{u(\al)v(\be)})_{\al\in A,\be\in B}$.
Let $f=\#\{\al\in A,u(\al)=\flat\}$
and ${\bar f}=\#\{\be\in B,v(\be)=\flat\}$
be the number of flat fields and antifields. The number of sharp fields and
antifields $s$, $\bar s$, and of natural ones $n$, $\bar n$ are defined
similarly.
For one of the determinants to be nonzero we must have $n={\bar n}$,
$f={\bar s}$ and $s={\bar f}$.
We can also group together fields and antifields according to their level
in the tree $\gT$, so that we get a block diagonal determinant that can be
factorized as:
$$
det(b_{\al\be}^{u(\al)v(\be)})_{\al\in A,\be\in B}
=\pm
\prod_{k=0}^{k_{max}}
det(b_{\al\be}^{\natural\natural})_{\al\in A_k^\natural,\be\in B_k^\natural}
$$
$$
\times
\prod_{k=0}^{k_{max}-1}
det(b_{\al\be}^{\sharp\flat})_{\al\in A_k^\sharp,\be\in B_{k+1}^\flat}
\prod_{k=1}^{k_{max}}
det(b_{\al\be}^{\flat\sharp})_{\al\in A_k^\flat,\be\in B_{k-1}^\sharp}
\eqno{(\rm III.4)}
$$
where $k_{max}$ is the maximal occupied level of the tree $\gT$.
$A_k^\natural$ is the number of $\al$'s in $A$ with $l^\gT(i(\al))=k$
and $u(\al)=\natural$. $A_k^\sharp$,
$A_k^\flat$,
$B_k^\natural$, $B_k^\sharp$ and
$B_k^\flat$
are defined in the same manner.
We must have of course $\#(A_k^\natural)=\#(B_k^\natural)$,
$\#(A_k^\sharp)=\#(B_{k+1}^\flat)$,
and $\#(A_k^\flat)=\#(B_{k-1}^\sharp)$ for any $k$, in order for the full
determinant to be nonzero.
Now by definition of $w_{ij}^{\gT,R}({\bf w})$ we have
$$
det(b_{\al\be}^{\natural\natural})_{\al\in A_k^\natural,\be\in B_k^\natural}
=
\det\biggl(C(x_{i(\al)},x_{j(\be)})
\de(c(\al),{\bar c}(\be))\biggr)_{\al\in A_k^\natural,\be\in B_k^\natural}
\ \ .\eqno{(\rm III.5)}
$$
Likewise
$$
det(b_{\al\be}^{\sharp\flat})_{\al\in A_k^\sharp,\be\in B_{k+1}^\flat}
$$
$$=
\prod_{\be\in B_{k+1}^\flat}w_{j(\be)j'(\be)}
\times
\det\biggl(C(x_{i(\al)},x_{j(\be)})
\de(c(\al),{\bar c}(\be))\biggr)_{\al\in A_k^\sharp,\be\in B_{k+1}^\flat}
\eqno{(\rm III.6)}
$$
where $j'(\be)$ denotes the ancestor of $j(\be)$ in the rooted tree $\gT$.
Finally
$$
det(b_{\al\be}^{\flat\sharp})_{\al\in A_k^\flat,\be\in B_{k-1}^\sharp}
$$
$$=
\prod_{\al\in A_{k}^\flat}w_{i(\al)i'(\al)}
\times
\det\biggl(C(x_{i(\al)},x_{j(\be)})
\de(c(\al),{\bar c}(\be))\biggr)_{\al\in A_k^\flat,\be\in B_{k-1}^\sharp}
\ \ .\eqno{(\rm III.7)}
$$
We see that the $w$'s factor out and are bounded by 1.
The remaining determinants again can be factorized according to the colors
$c((\al)$, and ${\bar c}(\be)$. Each determinant obtained that way is of
the form $det(C(x_{i(\al)},x_{j(\be)}))_{\al\be}$.

At this stage we can complete the proof of the Theorem
either using Gram's inequality or the method of comparisons of
rows and columns to bound these determinants.

Gram's inequality is the fastest road. We use (II.9)
to bound  every $p$ by $p$ determinant  by $K^{p}$.
And the complete loop determinant is then bounded by
$$
|det(b_{\al\be})_{\al\in A,\be\in B}|\le 3^{2n+2} K^{n+1}
\ \ .\eqno{(\rm III.8)}
$$
The end of the proof is as in the case of the Brydges-Kennedy interpolation.
\qed

It is also clear on the form (III.3) that the method of comparisons of
rows and columns of [IM], [FMRT] also works here. This method
roughly corresponds to Taylor expanding around
a middle point further and further when fields
or antifields accumulate in any given cube of unit size
of a lattice covering $\RR^{d}$.

\medskip
\noindent{\bf References}
\medskip

\item{[AR1]}{A. Abdesselam and V. Rivasseau, 
{\it Trees, forests and jungles: a
botanical garden for cluster expansions}, in Constructive Physics, 
Lecture Notes in Physics 446, Springer Verlag, 1995.}

\item{[AR2]}{A. Abdesselam and V. Rivasseau, 
{\it An Explicit Large Versus Small Field Multiscale
Cluster Expansion},  Rev. Math. Phys. Vol. 9 No 2, 123
(1997) (and references
therein).}

\item{[B]}{David Brydges, {\it Weak Perturbations of the Massless Gaussian
Measure}, in Constructive Physics, 
Lecture Notes in Physics 446, Springer Verlag, 1995 (and references
therein).}

\item{[BK]}{D. Brydges and T. Kennedy, {\it Mayer Expansions and the 
Hamilton-Jacobi Equation}, Journ. Stat. Phys. 48, 19 2064 (1978).
}

\item{[DR]}{M. Disertori and V. Rivasseau, in preparation.}

\item{[FKT]}{J. Feldman, H. Kn\"orrer and E. Trubowitz,
{\it A Representation for Fermionic Correlation Functions}, 
ETH-Z\"urich preprint June 1997.}

\item{[FKLT]}{J. Feldman, D. Lehmann, H. Kn\"orrer and E. Trubowitz,
{\it Fermi Liquids in Two Space Dimensions}, in Constructive Physics, 
Lecture Notes in Physics 446, Springer Verlag, 1995.}

\item{[FMRT]}{J. Feldman, J. Magnen, V. Rivasseau and E. Trubowitz,
{\it An Infinite Volume Expansion for Many Fermion Green's Functions}
Helv. Phys. Acta Vol. 65, 679 (1992).}

\item{[IM]}{D. Iagolnitzer and J. Magnen, 
{\it Asymptotic Completeness and Multiparticle
Structure in Field Theories, II, Theories with Renormalization},
Commun. Math. Phys. 111, 81 (1987).}

\item{[S]}{M. Salmhofer,  
Continuous renormalization for fermions and Fermi liquid theory,
cond-mat/9706188.}

\end